\documentclass[prl,reprint,twocolumn,superscriptaddress,showpacs, reprint,floats]{revtex4-1}
\usepackage{amsmath}
\usepackage{amsfonts}
\usepackage{amssymb}
\usepackage{euscript}
\usepackage{bm}
\usepackage[pdftex]{graphicx}

\usepackage{color}

\begin{document}

\title{Non-existence of the Luttinger-Ward functional and misleading
       convergence of skeleton diagrammatic series for Hubbard-like models}

\author{Evgeny Kozik}
\affiliation{Physics Department, King's College London, Strand, London WC2R 2LS, UK}
\affiliation{Centre de Physique Th\'eorique, Ecole Polytechnique, CNRS, 91128 Palaiseau Cedex, France}
\author{Michel Ferrero}
\affiliation{Centre de Physique Th\'eorique, Ecole Polytechnique, CNRS, 91128 Palaiseau Cedex, France}
\author{Antoine Georges}
\affiliation{Coll\`ege de France, 11 place Marcelin Berthelot, 75005 Paris, France}
\affiliation{Centre de Physique Th\'eorique, Ecole Polytechnique, CNRS, 91128 Palaiseau Cedex, France}
\affiliation{DPMC, Universit\'e de Gen\`eve, 24 quai Ernest Ansermet, CH-1211 Gen\`eve, Suisse}

\begin{abstract}
The Luttinger-Ward functional $\Phi[\mathbf{G}]$, which
expresses the thermodynamic grand potential in terms of the interacting single-particle Green's
function $\mathbf{G}$, is found to be ill-defined for fermionic models with the Hubbard on-site interaction. 
In particular, we show that the self-energy $\mathbf{\Sigma}[\mathbf{G}] \propto
\delta\Phi[\mathbf{G}]/\delta \mathbf{G}$ is \textit{not} a single-valued functional of $\mathbf{G}$:  
in addition to the physical solution for $\mathbf{\Sigma}[\mathbf{G}]$, there exists at least one qualitatively distinct unphysical branch. 
This result is demonstrated for several models: 
the Hubbard atom, the Anderson impurity model, and the full two-dimensional Hubbard model. 
Despite this pathology, the skeleton Feynman
diagrammatic series for $\mathbf{\Sigma}$ in terms of $\mathbf{G}$ is found to converge at least for moderately low temperatures. However, at strong interactions, its convergence is to the \textit{unphysical} branch. This reveals a new scenario of
breaking down of diagrammatic expansions. 
In contrast, the bare series in terms of
the non-interacting Green's function $\mathbf{G}_0$ converges to the correct physical branch of $\mathbf{\Sigma}$ in all cases currently accessible by diagrammatic Monte Carlo.
Besides their conceptual importance, these observations have important implications 
for techniques based on the explicit summation of diagrammatic series.  

\end{abstract}



\maketitle


The formalism of the Luttinger-Ward functional (LWF)~\cite{Luttinger_Ward} is a crucial constituent of the modern quantum many-body physics framework. Following Baym and Kadanoff~\cite{Baym_Kadanoff}, the free-energy is introduced as a functional of the full single-particle Green's function (GF) $\mathbf{G}$, which describes properties of single-particle excitations in a system of interacting particles~\footnote{Note that $\mathbf{G}$, $\mathbf{G}_0$ and $\mathbf{\Sigma}$ are understood as 
matrices, and that the trace in (\ref{Omega}) is over Matsubara frequencies (a prefactor 
$T=1/\beta$ is included in its definition) as well as all other relevant indices (momenta, internal indices, etc.).}:
\begin{equation}
\Omega[\mathbf{G}]\,=\,
\mathrm{tr} \ln \mathbf{G} 
- \mathrm{tr} \,[\mathbf{(G_0^{-1}-G^{-1})G}]+ \Phi[\mathbf{G}] \label{Omega}
\end{equation}
In this expression, $\mathbf{G_0}$ is the bare GF in the absence of interactions, and 
$\Phi[\mathbf{G}]$ is the LWF, which is the focus of the present paper. 
Defined only by the interaction term  
$H_{\mathrm{int}}$, the LWF is universal: the form of $\Phi[\mathbf{G}]$ does not depend explicitly on the quadratic part of the Hamiltonian (bare $\mathbf{G_0}$) and is shared by all systems with the
same structure of interactions between the particles.
In this formalism, the self-energy is also a functional of $\mathbf{G}$: 
\begin{equation}
  \mathbf{\Sigma}[\mathbf{G}]= \frac{1}{T} \frac{\delta \Phi[\mathbf{G}]}{\delta \mathbf{G}}, \label{Sigma}
\end{equation}
while the stationary point of $\Omega$, given by $\delta\Omega/\delta\mathbf{G}=0$, yields the 
Dyson equation $\mathbf{G}^{-1}-\mathbf{G_0}^{-1}+\mathbf{\Sigma}[\mathbf{G}]=0$, viewed 
here as a non-linear functional equation for $\mathbf{G}$ at equilibrium. 
 
The use of functionals $\Omega[\mathbf{G}]$, $\Phi[\mathbf{G}]$, $\mathbf{\Sigma[G]}$ has proven indispensable in a range of contexts from formal derivations to all orders in perturbation theory, such as that of Luttinger's theorem~\cite{Luttinger}, to devising approximations that are automatically consistent with sum-rules and conservation laws~\cite{Baym_Kadanoff}. Notable examples are the self-consistent Hartree-Fock approximation, or the 
dynamical mean-field theory (DMFT)~\cite{DMFT}. A number of extensions of DMFT have been recently proposed, formally based on the LWF, 
such as cluster (for reviews, see e.g.~\cite{maier_cluster_rmp_2005,kotliar_review_rmp_2006,tremblay_review_ltp_2006}) 
and diagrammatic~\cite{DiagMC_DMFT} extensions, 
the DMFT+GW method~\cite{biermann_gwdmft_prl,kotliar_nato_2002}, the dynamical
vertex approximation (D$\Gamma$A)~\cite{DGammaA}, DCA$^+$ \cite{DCA+}, etc.

\begin{figure}[t!]
\begin{center}
\includegraphics[width=1.0\columnwidth]{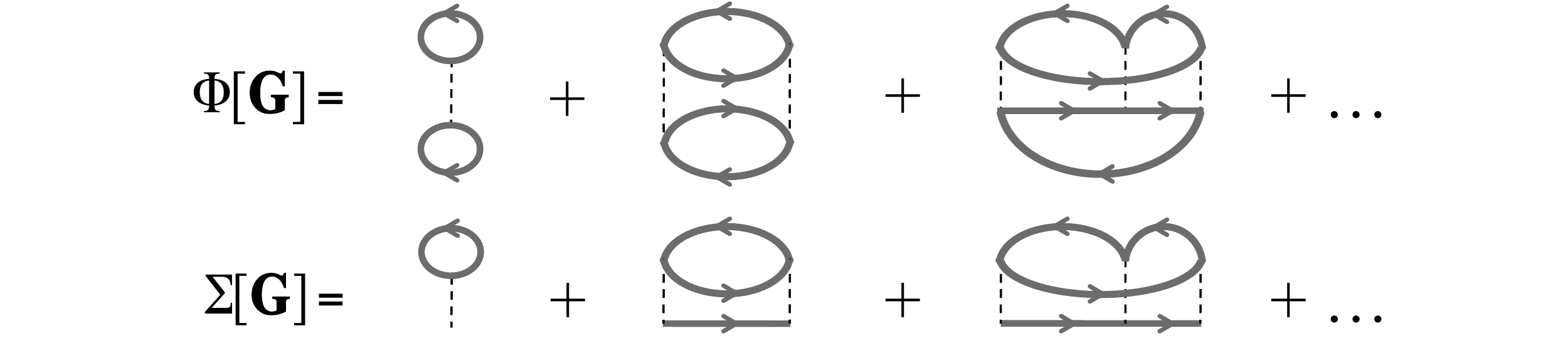}
\caption{Formal definition of $\Phi[\mathbf{G}]$ and $\mathbf{\Sigma}[\mathbf{G}]$ 
as skeleton diagrammatic expansions. The bold lines represent the full interacting GF $\mathbf{G}$ and the dashed lines the interaction vertex.}
\label{fig:diagrams}
\end{center}
\end{figure}

There are two ways to justify the formalism and propose a formal construction of 
the LWF $\Phi[\mathbf{G}]$, whose closed-form expression is unattainable in general. 
One is based on the diagrammatic perturbation expansion~\cite{Luttinger_Ward,AGD}, as illustrated 
in Fig.~\ref{fig:diagrams} for the Hubbard interaction  
$H_{\mathrm{int}}= U \sum_i  n_{i\uparrow} n_{i\downarrow}$. Here $\Phi[\mathbf{G}]$ is constructed explicitly as the sum of \textit{all} `skeleton' (also called `bold-line') diagrams in terms of $\mathbf{G}$, i.e. the diagrams that cannot be disconnected by cutting two propagator lines or, equivalently, 
that contain no self-energy insertions. 

The second approach, which is formally non-perturbative, is to view $\Phi$ as a Legendre-transform of 
the free-energy with respect to the single-particle GF~\cite{chitra_bk,georges_strong,potthoff_2003, potthoff_2006}
The idea is to constrain the GF to take a preassigned value $\mathbf{G}$ by appropriately choosing the bare propagator $\mathbf{G_0}$, thus viewed as a Lagrange multiplier. 
For the Legendre transform---and hence for $\Phi[\mathbf{G}]$---to be properly defined,  
the map $\mathbf{G_0}\rightarrow \mathbf{G}$ must therefore be {\it invertible}:  
there must be a \textit{unique} bare propagator $\mathbf{G_0}[\mathbf{G}]$ 
such that the interacting GF takes the value $\mathbf{G}$ for the specified Hamiltonian 
$H_{\mathrm{int}}$.

In this Letter, we show that the LWF for the Hubbard on-site interaction is ill-defined. In particular, $\mathbf{\Sigma}[\mathbf{G}]$ is found to be a non-single-valued functional of $\mathbf{G}$ with at least one qualitatively distinct unphysical branch, making the map $\mathbf{G_0}\rightarrow \mathbf{G}$ \textit{not} invertible. This is a non-perturbative statement, which does not rely on the diagrammatic definition of the LWF, and which
is based on evidences from several models with the Hubbard interaction. Although this finding does not necessarily undermine formal considerations based on the LWF, provided they are explicitly constrained to the physical branch, it has dramatic consequences for diagrammatic expansions.

The existence of two branches of $\mathbf{\Sigma}[\mathbf{G}]$ inevitably raises the question of what happens to the skeleton series for $\mathbf{\Sigma}$ in terms of $\mathbf{G}$, Fig.~\ref{fig:diagrams}. A natural scenario would be divergence in the spirit of Dyson's collapse argument~\cite{Dyson}. Indeed, it is believed that skeleton expansions are doomed to diverge at least at strong interactions or whenever the state of the system is not conformal to a Fermi liquid, like, e.g., a Mott insulator \cite{Hofstetter_Kehrein}. Here we demonstrate---by explicit summation using the diagrammatic Monte Carlo (DiagMC) technique~\cite{proceedings, FH}---that the skeleton series for $\mathbf{\Sigma}$ does converge even when $\mathbf{G}$ is that of a Mott insulator (at least at moderately low temperatures), the convergence with diagram order getting progressively faster at large $U$. However, at sufficiently strong interactions and close to half filling, the convergence is to the \textit{unphysical} branch, while in the weakly-correlated regime the skeleton series converges to the physical solution.
\footnote{ In fact, we first discovered that the skeleton expansion converges to an unphysical result and then traced
this back to the existence of two branches in the LWF.}

Consistently with the universality of the LWF, this qualitative conclusion applies to all models with the Hubbard interaction we considered, including the Hubbard atom and the single-site Anderson impurity model (for which the unphysical branch is independently found by non-perturbative means), as well as the two-dimensional Hubbard model~\cite{Hubbard}. 

Convergence of the skeleton diagrammatic series to an unphysical branch rather than its divergence in difficult regimes is a critical result for a wealth of analytic and numeric approaches, especially in light or the substantial recent interest in methods based on explicit summation of skeleton diagrams~\cite{fermipolaron, EOS, DiagMC_DMFT, spins}. It reveals a generic scenario of breaking down of skeleton expansions, in which there is no \textit{a priori} indication of the series becoming untrustworthy. We demonstrate, however, that, at least for the models considered here and in (unordered) regimes currently accessible by DiagMC, the \textit{bare} series in terms of $\mathbf{G}_0$~\footnote{Throughout the paper we include the mean-field Hartree contribution $\Sigma_\mathrm{H \, \uparrow, \downarrow}=U n_{\uparrow,\downarrow}$, where $n_{\uparrow,\downarrow}$ is the particle density per spin component, into $\mathbf{G}_0$ by appropriately shifting the chemical potential as explained, e.g., in Ref.~\cite{proceedings}} always converges to the \textit{physical} solution. This suggests that summations of bare diagrammatic series are intrinsically more reliable than those of skeleton expansions. 

Purely mathematically, even if a skeleton series converges, its convergence to the correct answer is not guaranteed. This is because it does not converge \textit{absolutely} \footnote{Here``absolutely'' implies that an absolute value is taken of each diagram.} at any coupling owing to the factorial number of terms in each order. For such a (conditionally convergent) series, the Riemann series theorem states that reordering its terms can make it converge to \textit{any} given number (or diverge). The skeleton series is, in fact, as a result of reordering of the corresponding bare series (see, e.g. \cite{AGD}). From this perspective, it is natural that the bare and the skeleton expansions converge to different answers, while the bare one is more robust being in essence the standard Taylor expansion in $U$. To our knowledge, our findings are the first observation of this possibility realized.

We start by addressing the existence of the functional $\mathbf{\Sigma[G]}$~\footnote{That this is different from assessing the number of solutions of $G$ for a given $G_0$, such as in GW,  where an explicit approximate form of $\Sigma[G]$ is used.
}.
Whenever a non-perturbative solution $\mathbf{G}[\mathbf{G_0}]$ is available for all possible $\mathbf{G_0}$, as e.g. for single-site models, the inverse relation $\mathbf{G}_0[\mathbf{G}]$ and hence $\mathbf{\Sigma}[\mathbf{G}] = \mathbf{G}_0[\mathbf{G}]^{-1}-\mathbf{G}^{-1}$ can be computed in practice 
using an iteration scheme. 
We thus attempt to compute
$\mathbf{\Sigma}[\mathbf{G}]$ for the simplest model with the Hubbard interaction, the Hubbard atom, $H_{\mathrm{at}}=U n_\uparrow n_\downarrow$, 
and for $\mathbf{G}$ equal to the exact solution $\mathbf{G}^{\mathrm{(exact)}} \equiv G^{\mathrm{(exact)}}(z) = \Big[ 1/(z + U/2) + 1/(z - U/2) \Big]/2$ 
(with $z=i\omega_n$ on the Matsubara axis). 
In this case, the physical self-energy and the bare GF are given by: $\Sigma^{\mathrm{(exact)}}(z)=U^2/4z$ and 
$G_0^{\mathrm{(exact)}}(z) = 1/z$. 
To this end we employ the following protocol: Starting from a guess $\mathbf{G}_0^{(n)}$, we find a certain $\mathbf{G}^{(n)}[\mathbf{G}_0^{(n)}]$
for the single-site problem using an interaction-expansion continuous-time quantum Monte Carlo
solver~\cite{solvers} implemented with the TRIQS~\cite{triqs} toolbox.
The next approximation for $\mathbf{G}_0$ is obtained with two schemes, A and B, given by
\begin{eqnarray}
[ {\mathbf{G_{0}^{-1}}}]^{(n+1)} =
[{\mathbf{G_0^{-1}}}]^{(n)} \pm
\Big( 
\mathbf{G}^\mathrm{(exact) \; -1} - [\mathbf{G^{-1}}]^{(n)} 
\Big).
\label{eq:scheme}
\end{eqnarray}
In scheme A, the $+$ sign is used for all Matsubara frequencies, while in
scheme B, $-$ is used for the lowest Matsubara frequency and $+$
for all the other ones. At convergence, both schemes coincide with Dyson's equation for the exact solution $\mathbf{G}^\mathrm{(exact)}$:
$\mathbf{G}_0^{-1} = \mathbf{G}^\mathrm{(exact) \; -1} + \mathbf{\Sigma}$; they are iterated until $\mathbf{G}$ matches $\mathbf{G}^\mathrm{(exact)}$ with arbitrary accuracy.
 
Strikingly, we observe that the schemes A and B converge to \textit{two} different solutions, the expected $\mathbf{G}_0^\mathrm{(exact)}[\mathbf{G}^\mathrm{(exact)}]$ and a drastically different $\mathbf{G}_0^\mathrm{(unphysical)}[\mathbf{G}^\mathrm{(exact)}]$. One can explicitly verify that both solutions satisfy the correct map $\mathbf{G}_0 \rightarrow \mathbf{G}$. 
This establishes, in a non-perturbative manner, that this 
map is not invertible, and that the functional $\mathbf{\Sigma}[\mathbf{G}]$ has at least two branches.  We emphasize that we have tried
other inverting methods, but all of them reproduced the results of either scheme A or B, suggesting that there are likely only two branches.

\begin{figure}[t!]
\begin{center}
\includegraphics[width=1.0\columnwidth]{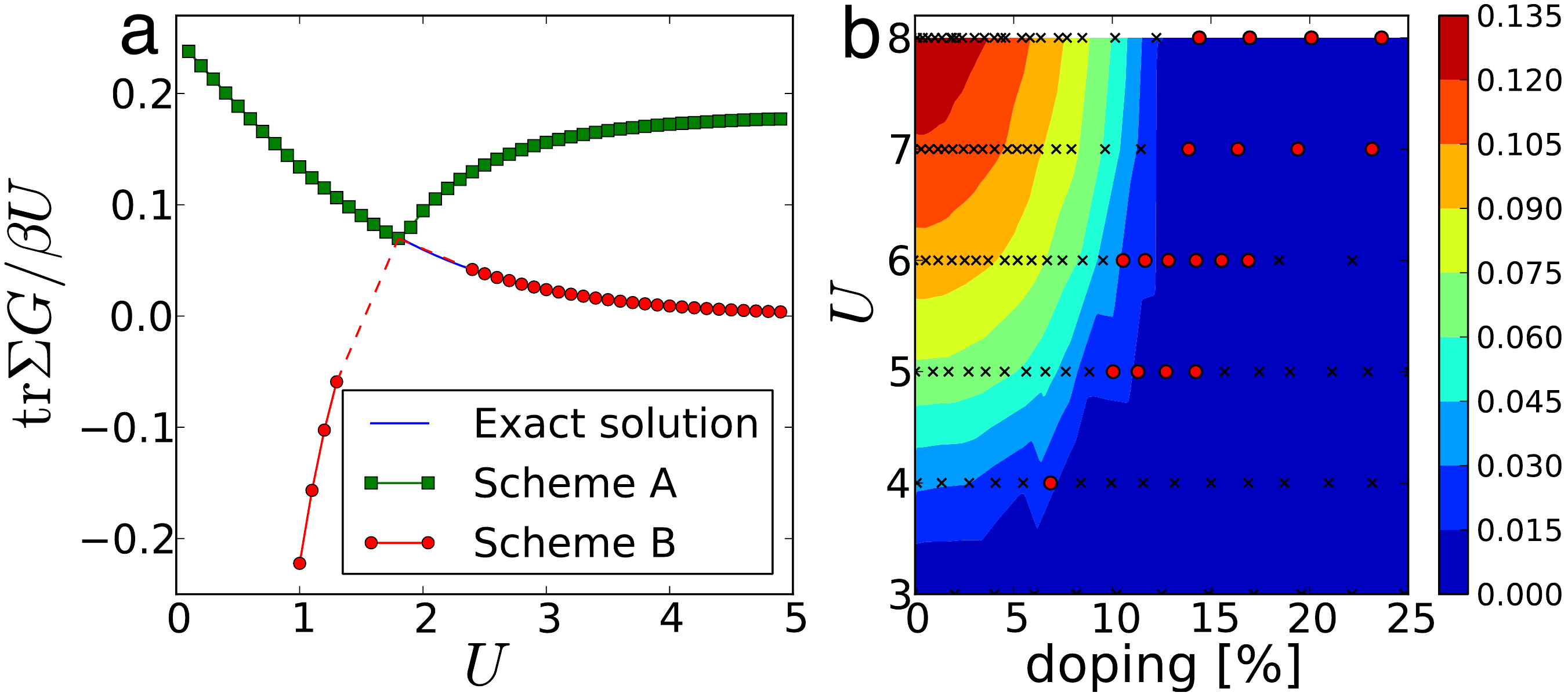}
\caption{(Color online) \textbf{(a)} Hubbard atom: Double occupancy vs. interaction at $T=0.5$, for the physical and unphysical branches (see text). \textbf{(b)} Anderson impurity: $|D^\mathrm{(exact)}-D^\mathrm{(found)}|$, quantifying the difference between the exact solution and that found by scheme A (see text),
in the $U$-$\delta$ plane at $T=0.5$. Black points (crosses) are converged calculations,
while at the red points (circles) the scheme A could not
converge.
\label{fig:D_Hubbard_Anderson_diagram}}
\end{center}
\end{figure}

The result is illustrated in Fig.~\ref{fig:D_Hubbard_Anderson_diagram}a [throughout the paper we use the units of the Hubbard model with the hopping $t=1$], showing the double occupancy $D=\langle n_\uparrow n_\downarrow \rangle=\mathrm{tr} \mathbf{\Sigma} \mathbf{G}^\mathrm{(exact)}/U$ for both solutions. The unphysical nature of the second solution is clearly seen in the corresponding $D$, which \textit{grows} with $U$. The two branches cross at a single value of the
interaction $U_*$. We observe that by using scheme A,
we follow the physical branch for $U<U_*$ and the
unphysical branch for $U>U_*$. Scheme B appears to have the opposite behavior, but it was impossible to converge
the results close to $U_*$. Moreover,
our results suggest that the unphysical branch could exist down to
small $U$ --- possibly even $U=0$ --- but the solution becomes increasingly
singular at $U \to 0$ and we could not converge our results for $U \lesssim 2$.

\begin{figure}[t!]
\includegraphics[width=1.0\columnwidth]{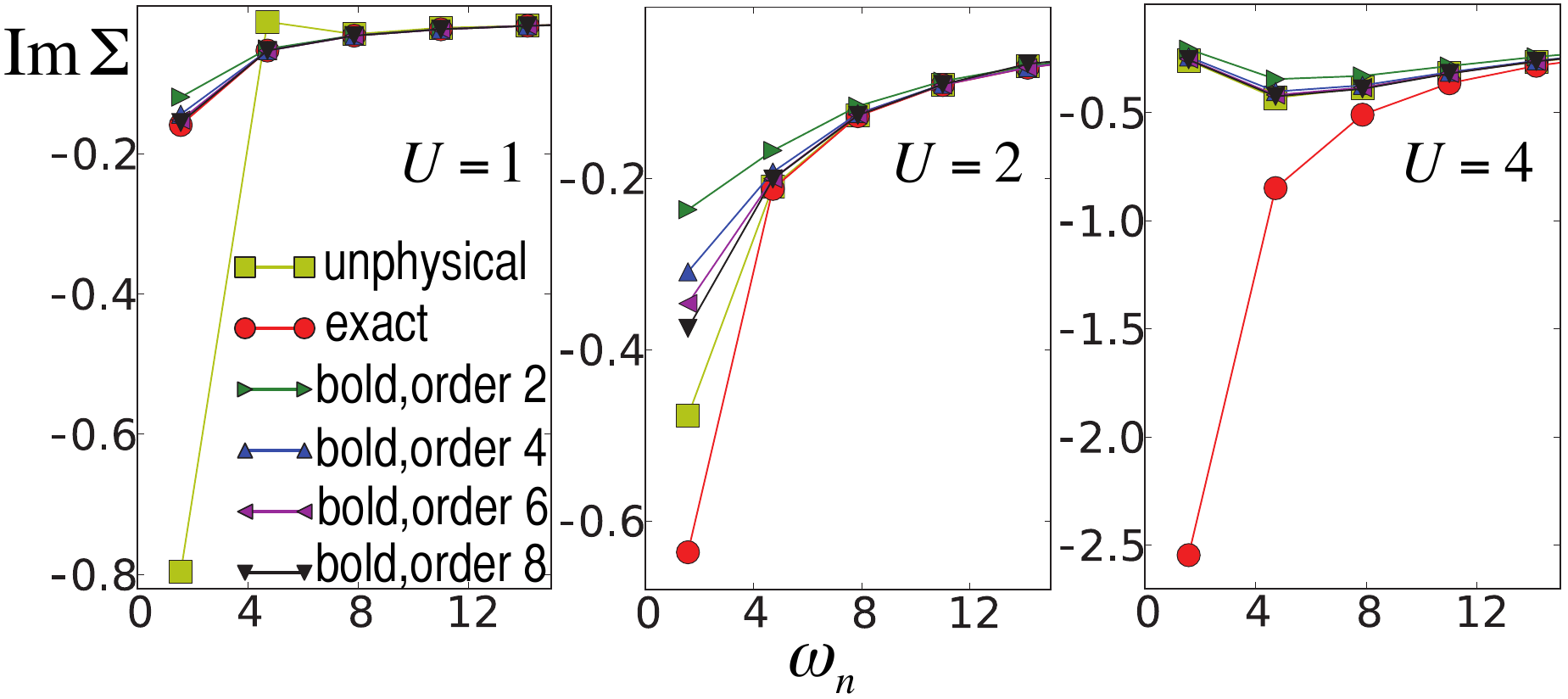}
\caption{(Color online) Hubbard atom: Two solutions for $\mathbf{\Sigma} [\mathbf{G}^\mathrm{(exact)}]$ and convergence of the corresponding skeleton (bold) expansion, Fig.~\ref{fig:diagrams}, at half filling, $T=0.5$ and various $U$.
}
\label{fig:al_diagmc}
\end{figure}
The Matsubara-frequency dependence of $\mathbf{\Sigma}$ for both solutions is compared in Fig.~\ref{fig:al_diagmc}. 
The perturbative high-frequency tails match. 
However, at $U>U_*$, $|\mathbf{\Sigma}^\mathrm{(unphysical)}|$  
becomes small at low frequencies instead of diverging as the exact solution does~\footnote{Since
both solutions correspond to
the same $\mathbf{G}^\mathrm{(exact)}$, the actual state of the system in both
cases is of course a gapped insulator.}. 
At $U<U_*$ $\Sigma^\mathrm{(unphysical)}$ is clearly pathological, in view, e.g., of the non-monotonicity at low
$i \omega_n$. 

It is worth noting that other pathologies and limitations of the LWF have been reported, e.g. in 
Refs.~\cite{georges_kotliar_comment, kiaran_phillips_kane_prl, Sangiovanni}. In the latter, it was found that 
$\delta \mathbf{\Sigma}[\mathbf{G}]/ \delta \mathbf{G}$ diverges at certain \textit{discrete} points in the parameter space. 
The immediate connection of this observation to our results is unclear and left for future work. 


We now turn to the question of what happens to the skeleton diagrammatic series for $\mathbf{\Sigma}[\mathbf{G}^\mathrm{(exact)}]$ in this case. DiagMC~\cite{proceedings,
FH} allows us to address it by a direct unbiased summation of the series to sufficiently high order. Partial sums of the series in terms of $\mathbf{G}^\mathrm{(exact)}$ for the Hubbard atom are plotted
in Fig.~\ref{fig:al_diagmc} up to order 8 (the highest accessible with our
computing resources). Despite the pathology of $\mathbf{\Sigma}[\mathbf{G}]$, the series appears convergent for all the values of $U$ we considered at least at moderately low temperatures. We identify three typical qualitative regimes: (i) for
$U<U_*$ the skeleton series clearly converges to the correct solution; (ii) for
$U \sim U_*$ the convergence becomes slow and it is unclear which solution the
series converges to since the two solutions are close in this regime, and (iii)
for $U>U_*$ the skeleton series exhibits fast convergence to the
\emph{unphysical} solution, the higher the value of $U$ the faster the convergence. 
Remarkably, for reasons so far unclear, this convergence behavior follows that of the iterative scheme A, Eq~(\ref{eq:scheme}). On the other hand, the bare series in terms of $\mathbf{G}_0^\mathrm{(exact)}$ always yields $\mathbf{\Sigma}^\mathrm{(exact)}$ for the Hubbard atom already at the second order---all the other diagrams exactly cancel in this case---which is a well-known analytic result. As an independent check of consistency of the map $\mathbf{G}_0 \rightarrow \mathbf{G}$, we have verified by DiagMC that the bare series in terms of $\mathbf{G}_0^\mathrm{(unphysical)}$ converges to $\mathbf{\Sigma}^\mathrm{(unphysical)}$.


Due to the universal nature of the LWF, these surprising results are not a unique feature of the Hubbard atom. We examined the single-impurity Anderson model with a conduction band described by a flat density of states on the interval $[-1,1]$ and an energy-independent hybridization $V=1$ for different values of the interaction $U$ and the doping $\delta$. We first find the numerically exact $\mathbf{G}^\mathrm{(exact)}$ for a given set of parameters using the impurity solver.
This GF is then used in the iterative
scheme A.
Fig.~\ref{fig:D_Hubbard_Anderson_diagram}b reports the difference $|D^\mathrm{(exact)}-D^\mathrm{(found)}|$ which quantifies the remoteness of the found solution from the exact
solution in terms of the double occupancy. The manifold in the space of parameters where $|D^\mathrm{(exact)}-D^\mathrm{(found)}|$ disappears corresponds to the intersection of the two branches, dividing the parameter space into two qualitatively distinct regions: at larger values of $U$ around $\delta=0$ scheme A converges to the unphysical solution (with a
critical doping increasing with $U$), while it converges to the correct solution at small $U$ and large $\delta$. We note that near the boundary between the regimes the convergence of scheme A becomes extremely slow. We have also performed DiagMC calculations for 
several points on Fig.~\ref{fig:D_Hubbard_Anderson_diagram}b, showing that the skeleton series in terms of $\mathbf{G}^\mathrm{(exact)}$ always converges to the same solution as the iterative scheme A, except close to the boundary, where the diagram convergence also becomes slow, adding to the remarkable conspiracy. Hence, Fig.~\ref{fig:D_Hubbard_Anderson_diagram}b can be viewed as the convergence diagram for the skeleton series. The qualitative shape of the diagram is expected to be shared by all the models with the Hubbard interaction in view of the universality of the LWF.

Our DiagMC simulations of the \textit{bare} series for the self-energy in terms of $\mathbf{G}_0^\mathrm{(exact)}$ show that, whenever we can access sufficiently high orders to reach convergence, the bare series converges to the \emph{exact} solution $\mathbf{\Sigma}^\mathrm{(exact)}$ in both regions of the diagram.


\begin{figure}[t!]
\begin{center}
\includegraphics[width=1.0\columnwidth]{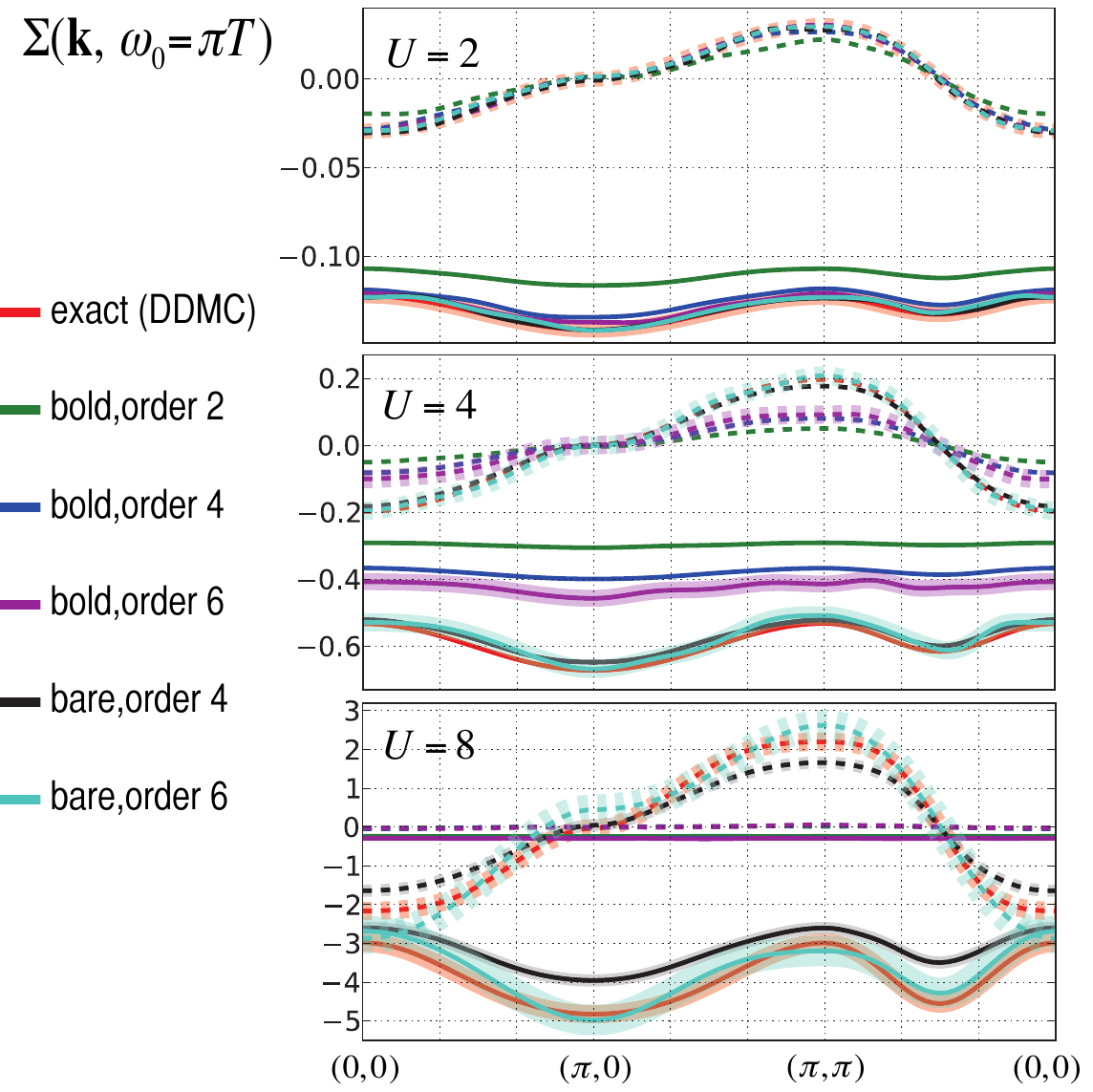}
\caption{(Color online) 2D Hubbard model: Convergence of the skeleton (bold) and bare
series for the momentum dependence of $\mathbf{\Sigma}$ at the lowest Matsubara
frequency at half filling, $T=0.5$. The solid (dashed) lines are $\mathrm{Re} \mathbf{\Sigma}$ ($\mathrm{Im} \mathbf{\Sigma}$), the widths express the corresponding error bars.}
\label{fig:Sigma_k_Hubbard}
\end{center}
\end{figure}

We complete our study with the most interesting case of the Hubbard model on the square lattice, 
for which the
self-energy acquires non-trivial momentum dependence. At half-filling, reliable
benchmarks are available from unbiased diagrammatic determinant Monte Carlo
(DDMC) calculations~\cite{DDMC}. We use the full GF $\mathbf{G}^\mathrm{(exact)}$ obtained by
DDMC as an input for the DiagMC summation of the skeleton series for
$\mathbf{\Sigma}$. In parallel, we employ DiagMC to sum the
corresponding bare series. The results are compared in
Fig.~\ref{fig:Sigma_k_Hubbard}, showing the momentum dependence of
$\mathbf{\Sigma}$ at the lowest Matsubara frequency $\omega_0=\pi T$ at fixed $T=0.5$ and various interaction strengths.  The qualitative behavior
is identical to that observed in the single-site models. We see that
the \textit{bare} series reproduces the DDMC benchmark for $\Sigma(\mathbf{k},
\omega_0)$ within the error bars for all the interaction strengths considered
(admittedly, convergence as a function of diagram order becomes slower at
larger $U$). On the contrary, the \textit{skeleton} series reliably converges
to the correct solution only at $U=2$, while at $U=8$ it displays fast
convergence to an almost momentum-independent function, drastically different
from the exact solution. At the intermediate $U=4$ the convergence of the
skeleton series becomes slow, very similarly to the case in the second panel of
Fig.~\ref{fig:al_diagmc}, suggesting that the value of $U$ is close to the
crossing point $U_*$ between the two branches. Unlike the single-site case,
we have no means of accessing the unphysical branch of
$\mathbf{\Sigma}[\mathbf{G}]$ other than by summing the skeleton series
explicitly.


Interestingly, the map $\mathbf{G_0}\rightarrow \mathbf{G}$ is
known~\cite{potthoff_2003} to be invertible if $\mathbf{G_0}$ is constrained to
the form $[\mathbf{G}^{-1}_0]_{ij}=i\omega_n+\mu-t_{ij}$ (with $i,j$ the lattice sites).
Consistently, our $\mathbf{G}_0^\mathrm{(unphysical)}$ contains an additional
frequency-dependent hybridization $\mathbf{\Delta}$,
$[\mathbf{G}_0^{\mathrm{(unphysical)}
-1}]_{ij}=i\omega_n+\mu-t_{ij}-\mathbf{\Delta}_{ij}(i\omega_n)$.
It is unclear, however, how this can be used to render the skeleton series convergent to $\mathbf{\Sigma}^\mathrm{(exact)}$ since $\mathbf{\Sigma}[\mathbf{G}]$ has no explicit dependence on $\mathbf{G}_0$.
In practical diagrammatic calculations~\cite{AGD}, when
$\mathbf{G}^\mathrm{(exact)}$ is unknown, $\mathbf{G}$ is found by
(iteratively) solving the Dyson equation
$\mathbf{G}^{-1}=\mathbf{G}_0^{-1}-\mathbf{\Sigma}[\mathbf{G}]$ with
\textit{true} $\mathbf{G}_0 \equiv \mathbf{G}_0^\mathrm{(exact)}$.
Clearly, in the regimes where the series for $\mathbf{\Sigma}[\mathbf{G}^\mathrm{(exact)}]$ converges to the unphysical branch, $\mathbf{\Sigma}[\mathbf{G}^\mathrm{(exact)}] \neq \mathbf{\Sigma}^\mathrm{(exact)}$, the calculation \textit{cannot} yield the correct answer $\mathbf{G}^\mathrm{(exact)}$. Provided there are no obvious pathologies in the unphysical $\mathbf{\Sigma}$,
which is the case in the examples considered here, identifying that the obtained $\mathbf{G}$ is wrong may be practically impossible in some computation schemes without an
\textit{a priori} benchmark.



To summarize, we have demonstrated that the LWF for the Hubbard interaction has at least two
branches, possibly everywhere in parameter space. The branches cross along a manifold, dividing the space of parameters into the ``weakly-correlated'' region, where the skeleton series converges to the physical solution, and the ``strongly-correlated'' region, where the skeleton series converges to the unphysical branch (as qualitatively described by
Fig.~\ref{fig:D_Hubbard_Anderson_diagram}b)~\footnote{Strictly speaking there must be a third region at sufficiently low temperatures where diagrammatic series must break down due a phase transition.}. We emphasize that the strongly-correlated
region does not need to be the insulating regime of the system---the skeleton series for the Anderson model can converge to the unphysical branch even in the correlated \textit{metallic} state. The boundary
between these regions is characterized by an increasingly slow convergence (possibly divergence) of the skeleton series. In contrast, we have found that the \textit{bare} series in terms of the non-interacting GF
$\mathbf{G}_0$ is insensitive to this boundary and converges to the
physical branch in both regions, although stronger interactions require
increasingly high orders to claim the converged result. At large interactions, it eventually becomes impossible to reliably extrapolate to the
infinite-order limit, and new developments are needed to be able to reach higher expansion orders.

\acknowledgments{
We acknowledge support of the Swiss National Science Foundation (Fellowship for Advanced Researchers), the DARPA/MURI OLE program, 
the European Research Council (under grant ERC-319286 QMAC) and the Simons Foundation (Many Electron Collaboration program).  
We acknowledge useful discussions with E.~Burovski, Y.~Deng, A.~J.~Millis, N.~Prokof'ev, G.~Sangiovanni, B.~Svistunov, K.~van~Houcke and F.~Werner. }


\bibliography{bibliography}

\end{document}